\documentclass[11pt]{article}

\usepackage{mathptmx}       
\usepackage{helvet}         
\usepackage{courier}        
\usepackage{type1cm}        
\usepackage{graphicx}        
\usepackage[bottom]{footmisc} 
\renewcommand{\bibitem}{}
\newcommand{\bhl}{}
\newcommand{\ehl}{}

\newcommand{\apj}{\textit {Astrophys.~J. }}%
\newcommand{\apjl}{\textit {Astrophys.~J. }}%
\newcommand{\aap}{\textit {Astron. Astrophys. }}%
\newcommand{\mnras}{\textit {Mon. Not. Roy. Astron. Soc. }}%
\newcommand{\pre}{\textit {Phys.~Rev.~E }}%
\newcommand{\pasj}{\textit {Pub. Astron. Soc. Japan }}%
\newcommand{\solphys}{\textit {Sol.~Phys. }}%
\begin{document}

\title{Theories of the solar cycle : a critical view}
\author{ H.C.\  Spruit\footnote{\tt henk@mpa-garching.mpg.de}\\Max Planck Institute for Astrophysics\\Box 1317, 85741 Garching, Germany }
\date{}
%
%
\maketitle

\begin{abstract}Some established views of the solar magnetic cycle are discussed critically, with focus on two aspects at the core of most models:  the role of convective turbulence, and the role of the `tachocline' at the base of the convection zone. The standard view which treats the solar cycle as a manifestation of the interaction between convection and magnetic fields is shown to be misplaced. The main ingredient of the solar cycle, apart from differential rotation, is instead buoyant instability of the magnetic field itself. This view of the physics of the solar cycle was already established in the 1950s, but has been eclipsed mathematically by mean field turbulence formalisms which make poor contact with observations and have serious theoretical problems. The history of this development in the literature is discussed critically.  The source of the magnetic field of the solar cycle is currently assumed to be located in the `tachocline': the shear zone at the base of the convection zone. While the azimuthal field of the cycle is indeed most likely located at the base of the convection zone, it cannot be powered by the radial shear of the tachocline as assumed in these models, since the radiative interior does not support significant shear stresses. Instead, it must be the powered by the latitudinal gradient in rotation rate in the convection zone, as in early models of the solar cycle. Possible future directions for research are briefly discussed.
\end{abstract}

\section{The role of convective turbulence}

For a star to generate a self-sustained magnetic field, it is sufficient that it rotate differentially. This differs from the traditional view of dynamos in stars, which holds that in addition to the shear flow due to differential rotation, a small scale velocity field has to be imposed in order to `close the dynamo cycle', thus enabling a selfsustained field independent of initial conditions. Convection can provide such a velocity field, and in fact convection has become such an integral part of thinking about dynamos in stars that the subject of `stellar magnetic fields' has been almost synonymous with `convective dynamos' for decades (for reviews see e.g. Weiss 1981 - 1997, R\"udiger and Hollerbach 2004, Tobias 2005, for recent texts Brandenburg 2009,  Jones et. al. 2009, \bhl Charbonneau 2005\ehl). Whether or not such a dynamo process can take place in principle is a separate matter. From the observations it is evident, however,  that it is not the way the solar cycle works. Instead, as I will argue below, the cycle operates on dynamic instability of the magnetic field itself. Convection plays only an indirect role, namely by maintaining the differential rotation of the envelope from which the cycle derives its energy.

\subsection{Mechanism of the solar cycle as inferred from observations}
\label{obs}

The common ingredient in all dynamo models such as those for the Earth's magnetic field or the solar cycle is the generation of a toroidal (azimuthally directed) field by stretching (`winding-up') of the lines of a poloidal field (e.g. Elsasser 1956). This is `the easy part'. It produces a field that increases in strength linearly with time and is proportional to the imposed initial field. To produce a cyclic, self-sustained field as observed there must be a second step that turns some of the toroidal field into a new poloidal component, which is again wound up, completing a field-amplification cycle that becomes independent of initial conditions. The particular process by which the new poloidal field is generated distinguishes the models from each other. In early models of the solar cycle that were popular in their time (Babcock 1961, 1963, Leighton 1969) observations of the emergence of active regions were used to infer the nature of the process responsible for this key step in the dynamo cycle. 

\begin{figure}
\hfil \includegraphics[width=0.6\linewidth, clip]{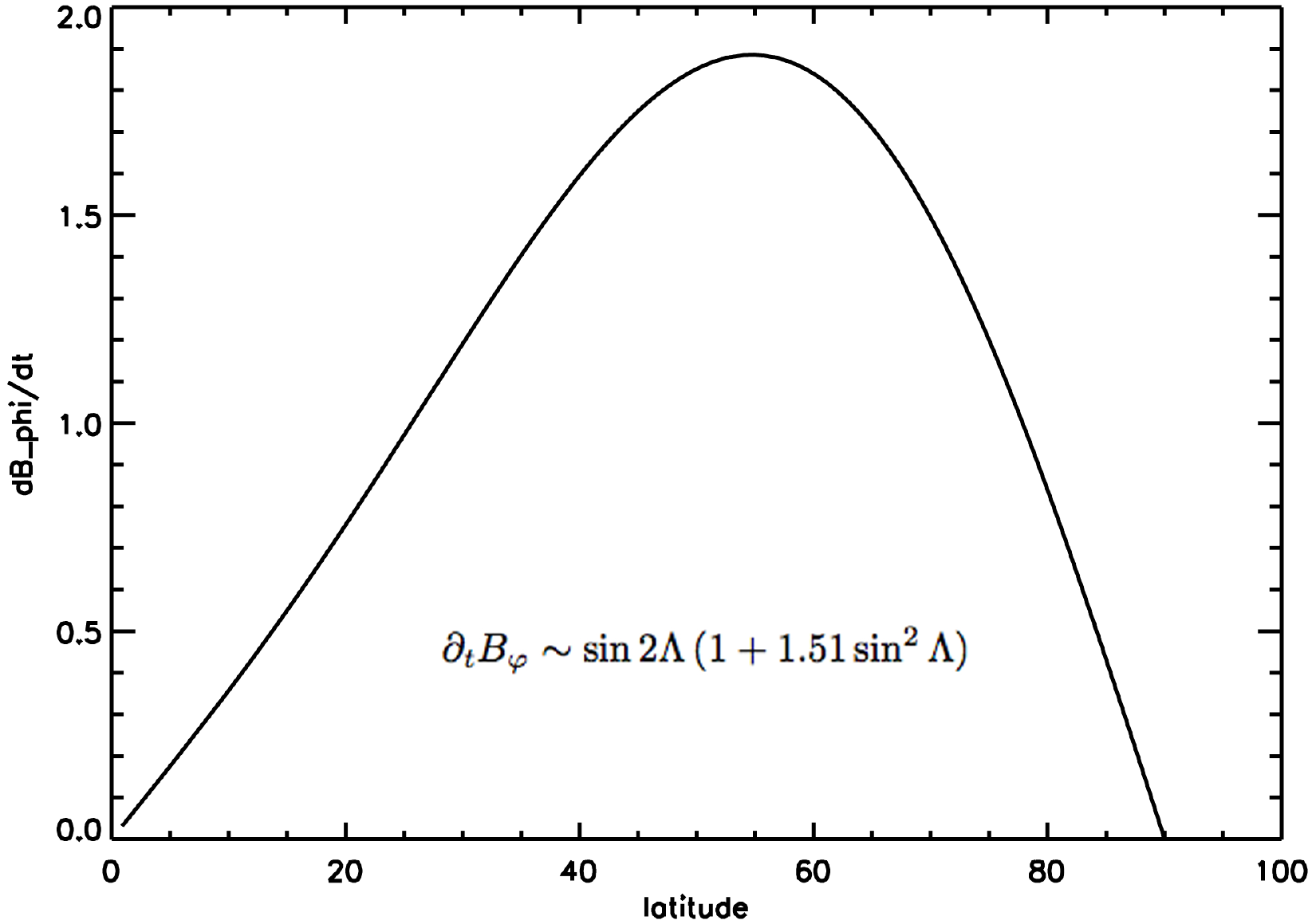}\hfil
\caption{\small Rate of increase of the azimuthal field strength as  a function of heliographic latitude, due to the observed differential rotation acting on an assumed uniform poloidal field.}
\label{dbdt}
\end{figure}

\begin{figure}
\hfil\includegraphics[width=0.9\linewidth, clip]{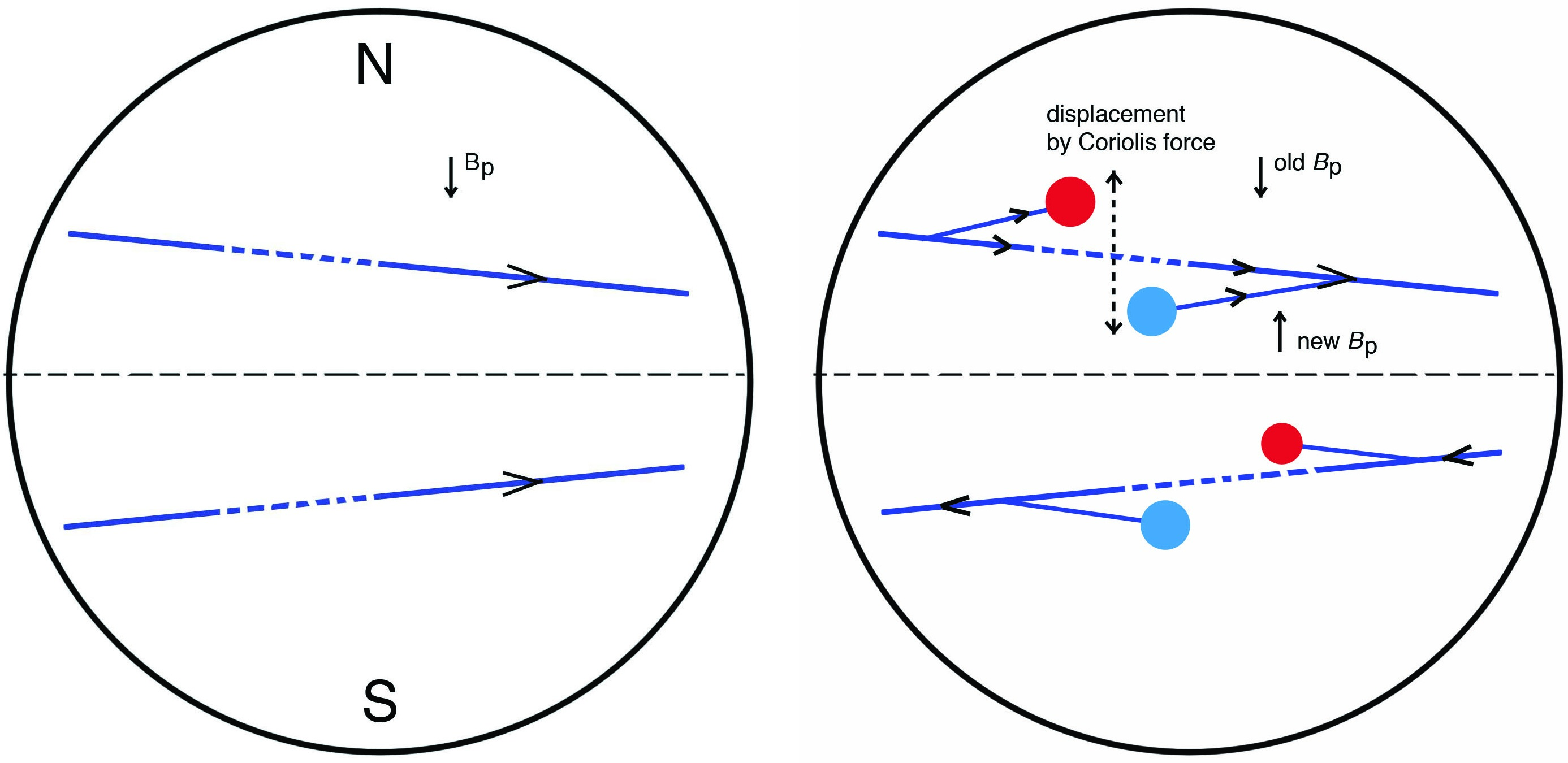}\hfil
\caption{\small Closing of the dynamo cycle by active region emergence. Left: sub-surface field produced by stretching of a poloidal field $ {\bf B}_{\rm p}$ by differential rotation (equator rotates faster). Coriolis forces during emergence of a stretch of the field (broken) to the surface causes displacements of the footpoints, observed at the surface as `tilt' of the active regions (circles). At depth, this produces a new poloidal field component of opposite sign.}
\label{sketchemerg}
\end{figure}

\begin{figure}[h]
\hfil\includegraphics[width=0.8\linewidth, clip]{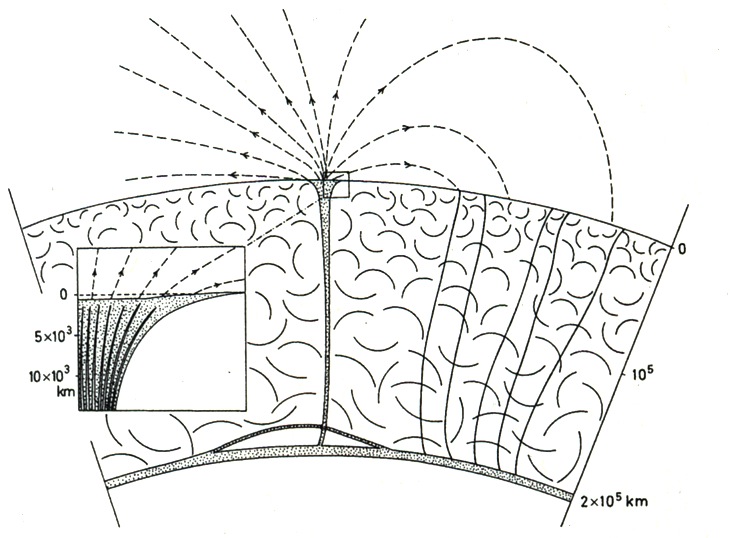}\hfil
\caption{\small Vertical cut through an active region illustrating the connection between a sunspot at the surface and its origins in the toroidal field layer at the base of the convection zone. (From Spruit and Roberts 1983).}
\label{spotsketch}
\end{figure}

These models proposed that the increasing toroidal field eventually becomes unstable, erupting to the surface to form the observed active regions (Cowling 1953, Elsasser 1956, Babcock 1961, see sketches in Figs.\ \ref{sketchemerg}, \ref{spotsketch}). The equatorward drift of the main zone of activity reflects the latitude dependence of the time taken for the toroidal field to reach the point of buoyant instability (Babcock 1961, 1963). This is illustrated with a simple model in Fig.\ \ref{dbdt}. In this sketch, a uniform poloidal field is assumed to be stretched  passively by the latitudinal differential rotation as observed on the surface of the Sun. Helioseismic observations (see review by Howe, 2009) show that this pattern of rotation also holds to a fair approximation inside the convection zone.  

The azimuthal field becomes unstable to buoyant rise when a critical strength of $\sim 10^5$ G is reached (Sch\"ussler et al. 1994). This happens first at the latitude where the rate of increase of the field is largest, around a latitude of $60^\circ$ in the simple model of Fig.\ \ref{dbdt}. This agrees with observations (Altrock 2010), though initially only small-scale magnetic activity without sunspots is produced. As time progresses, the field also becomes unstable at lower latitudes, producing an equatorward drift of the zone of activity. For reasons unknown, sunspots form only below a latitude of around 40$^\circ$. As Fig.\ \ref{dbdt} implies, Babcock's model also predicts a poleward propagating branch. Such a branch (but without sunspots)  is actually present on the Sun (the `poleward rush', Leroy \& Trellis, 1974, Altrock 2010). Its observational status and interpretation are not entirely clear, however.

\begin{figure}
\includegraphics[width=0.24\linewidth, clip]{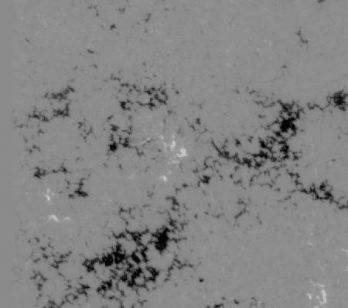}\hfil\includegraphics[width=0.24\linewidth, clip]{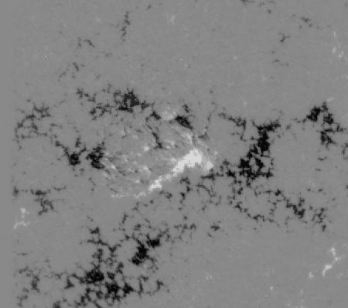}\hfil\includegraphics[width=0.24\linewidth, clip]{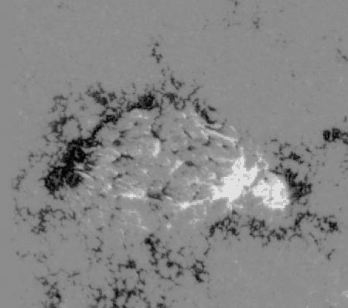}\hfil\includegraphics[width=0.24\linewidth, clip]{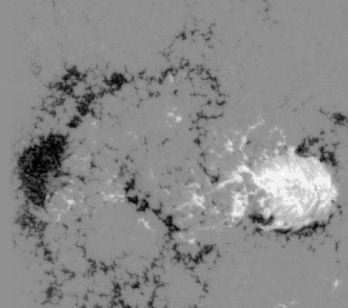}
\caption{\small Sequence (time from left to right) showing the emergence of an active region at the solar surface observed with the Hinode satellite. The opposite magnetic polarities (vertical component of the field) are shown in black and white. For a movie of this sequence see {http://science.nasa.gov/headlines/y2007/images/trilobite/Hinode\_lower.mov} }
\label{trilobite}
\end{figure}

The process of emergence of an active region has been studied in great detail for more than a century. A small patch of fragmented magnetic fields with mixed polarities appears and expands as more flux emerges (Fig.\ \ref{trilobite}). The surroundings of this patch remain unaffected by this process. The mix of polarities then separates into two clumps, the polarities traveling in opposite directions to their destination, ignoring the convective flows in the region. 

\begin{figure}[h]
\hfil\includegraphics[width=0.5\linewidth, clip]{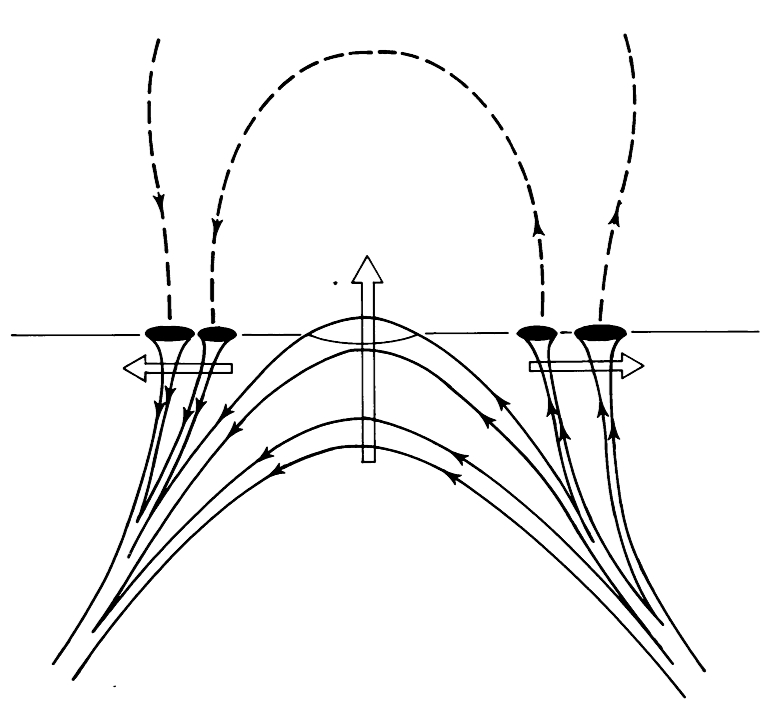}\hfil
\caption{\small `Rising tree' sketch to explain the phenomenology of a active region emergence. (From Zwaan, 1978)}
\label{zwaan}
\end{figure}

This striking behavior is the opposite of diffusion. To force it into a diffusion picture, one would have to reverse the arrow of time. Instead of opposite polarities decaying by diffusing into each other, they segregate out from a mix. The MHD equations are completely symmetric with respect to the sign of the magnetic field, however. There are no flows (no matter how complex) that can separate fields of different signs out of a mixture. \bhl This rules out a priori all models attempting to explain the formation of sunspots and active regions by turbulent diffusion. For recent such attempts, which actually ignore the observations they are trying to explain, see Kitiashvili  et al. (2010), Brandenburg et al. (2010).  The observations, instead, demonstrate that the orientation and location of the polarities seen in an active region must already be have been present in the initial conditions: in the layers \ehl below the surface from which the magnetic field traveled to the surface.  

The fragmented state near the surface in the early stages of the eruption process is only temporary. The intuitive `rising tree' picture (Zwaan, 1978) illustrates this (Fig.\ \ref{zwaan}). The observed fragmentation and subsequent formation of spots from a horizontal strand of magnetic field below the surface has recently been reproduced in striking realism in full 3-D radiative MHD simulations (Cheung et al. 2008, Rempel, this volume).

The axes of active regions are observed to be tilted with respect to the east-west direction. This was  attributed to the action of Coriolis forces during the emergnece process, and identified with the generation of the new poloidal field component that closes the dynamo cycle by Leighton (1969, see sketch in Fig.\ \ref{sketchemerg}). 

\subsection{Later developments}

Models like Leighton's thus made a direct connection between observations of the active regions that make up the solar cycle and the functioning of the cycle as a whole. One might have expected that this natural state of affairs would have led to a further development of the theoretical ideas in continued contact with the observations. But this has not been the case. 

Instead, the development of these ideas has been eclipsed for several decades by the parallel development of turbulent mean field formalisms for the solar cycle. These ideas postulated mathematically tractable equations which were claimed to represent the physics of the interaction between magnetic fields and convection in some statistical sense. They relied on theoretical assumptions like cascades in wavenumber space, correlation functions to represent the interaction between magnetic fields and flows, and an assumed separation of length scales between mean fields and fluctuations. Just looking at the data as described above, it is difficult see how a separation would be accomplished. What is more, the data themselves already contain more detailed and more critical information on the functioning of the cycle than is present in mean field models.  The dominance of these formalisms in the astrophysical literature (thousands of papers) has led to a particularly sterile theoretical view of the solar cycle, supported neither by a sound theoretical foundation of the equations used nor making much contact with the observations. 

In addition, it has had the effect of obscuring an important fact, namely that no turbulence needs to be imposed at all for dynamo action to take place.  A system that is completely laminar in the absence of magnetic fields can produce dynamo action from shear and magnetic instability alone (cf. Spruit 2002). A well studied and very successful example of such a dynamo process is the MRI turbulence observed in numerical simulations of accretion disks (e.g. Hawley et al. 1996). The models by Babcock and Leighton are just  another example where magnetic instability is the key element in closing the field amplification cycle. These kinds of magnetic cycle are intrinsically non-linear (i.e. not `kinematic' in dynamo parlance): their functioning depends on the finite amplitude of the field generated. This is because the time scale of the magnetic instabilities that close the dynamo cycle depends on field strength.

The conditions for self-sustained field generation to occur by differential rotation and instability alone, the properties of the magnetic field produced in this way, and its observable consequences all reflect the nature of the magnetic instability involved. \bhl In the case of the solar cycle: the properties of magnetic buoyancy. It is sometimes argued that such a process just  brings about an `alpha effect', so that one just has to  use \ehl a set of equations that incorporate such an effect. Neither the fact that a poloidal field component can appear by a process changing the direction of an initially toroidal field, however, nor the fact that turbulent mean field equations contain a term describing such an effect, are justifications for using these equations. An understanding of the solar cycle, or any other dynamo process, requires physics to be found out first, rather than assumed in some parametrized form. The idea that insight about the solar cycle can be obtained from the solutions of such models has been an impediment to real progress, however tempting the equations may have looked.

Justification for this critical view is found in the history of ideas about the solar cycle; this is done in the following section. I briefly discuss there how mean field thinking has led to a systematic disconnect between theory and observations. In all likelihood this would not have been necessary if the observations and their interpretation in models such as Leighton's (1969), had been taken more serious. 

\section{Failure of convective dynamos models of the solar cycle}

The turbulent view of magnetic field generation in convective stellar envelopes holds that the generation of a new poloidal field from the toroidal field produced by differential rotation should be seen as due to the effect of convection acting on a magnetic field in a rotating fluid (Parker 1955, 1979, \bhl Steenbeck et al. 1966\ehl,  Weiss 1981). A consequence of this model is therefore that dynamo action takes place throughout the convection zone. In this model the equatorward drift of the main belt of activity during the cycle reflected the (at the time unknown) {\em radial} gradient in the rotation rate, not the observed latitudinal differential rotation that were key to the Babcock and Leighton models. This is because mean field dynamo equations naturally lead to dynamo waves traveling in a direction perpendicular to surfaces of constant rotation rate. The observed equatorward drift of activity during the cycle therefore required the rotation rate to have a predominantly radial gradient, with rotation increasing inward.  Finally, it was noted that the Lorentz force limits the action of convection on magnetic fields when the field strength reaches equipartition with the kinetic energy of the convective `eddies' (e.g. Proctor and Weiss 1982). This predicts that the field in the solar envelope should be an intermittent turbulent field, with intrinsic field strengths of a few thousand Gauss. 

\subsection{Predictions}

Mean field dynamo models thus made three testable predictions. (They are found in many of the texts of the 70's and 80's, where they appear mostly as accepted consensus rather than as testable predictions):

\begin{itemize}
\item{} The dynamo action takes place through interaction with turbulent convection,
\item{} The rotation rate in the convection zone depends mainly on radius, it increases with depth.
\item{} The field strength does not exceed equipartition with convective energy densities (few thousand Gauss). 
\end{itemize}

These predictions have never agreed with the phenomenology of the solar cycle very well. One of the important observations is Hale's polarity law: the fact that magnetic fields are not present on the surface in a random `turbulent'  form, but appear in a strikingly systematic way, as bipolar active regions oriented east-west, with one of the polarities systematically leading (in the direction of rotation).  On top of the east-west orientation, the leading polarity is systematically shifted towards the equator compared with the following polarity (Joy's law, see sketch in Fig.\ \ref{sketchemerg}). To circumvent these observations, dynamo theories \bhl ignore heliographic longitude (of sunspot locations\ehl, for example); parameters of the model are then adjusted to fit the remaining data (the `butterfly diagram'). Given the reduced nature of these data and the degrees of freedom of the models, this process is usually successful. The price paid is that most observations of active region phenomenology have to be declared irrelevant, when in fact they provide the most telling evidence about the operation of the cycle. This attitude has remained an integral part of mainstream dynamo thinking. 

 Attempts have been made to reconcile the magnetic eruption view of active regions with the role of convective turbulence \bhl in the mean field dynamo view\ehl. Weiss (1964) proposed that magnetic fields rise from the interior, but that turbulence takes over in bringing about observations like the formation of sunspots. In this view, sunspots would form by random walk of magnetic fields in a turbulent flow. This proposal thus kept Cowling's view of active regions as emerging from below, but effectively discarded the observational evidence that led to this idea in the first place, namely the formation of sunspots. Meyer et al (1974) explicitly repeated the view that sunspots form by random walk of magnetic field lines in convective turbulence. 

This has been challenged by observers, who noted that sunspots do not form randomly but in a strikingly deterministic way, as described above (e.g. Zwaan 1978). The motion of active region magnetic fields independent of and opposing surface flows is documented by virtually all observations of active region formation (e.g. Tarbell et al. 1990, \bhl Strous et al. 1996\ehl). The consequence, namely that the magnetic field itself, rather than convective turbulence, forms active regions has always contradicted the role of convection assumed in mean field models. 

Since then, helioseismic measurements of the internal rotation have shown the second prediction to be wrong as well: the differential rotation is in fact mainly in latitude. The radial gradient is weak, and where it is present it is mostly of the wrong sign. 


\bhl The third prediction can be tested somewhat more indirectly by making use of the many clues given by observations of active regions.  A major step forward in the interpretation of this phenomenology \ehl are the simulations of flux bundles rising from the base of the convection  by Moreno-Insertis (1986), D'Silva and Choudhuri (1993), Fan et al. (1994), Sch\"ussler et al. (1994). In these simulations, a horizontal (azimuthal, zonal) bundle of magnetic field lines at the base of the convection zone is allowed to become unstable and rise to the surface. The degrees of freedom in these calculations are the magnetic flux of the bundle (set by the value observed for a typical active region), and its initial field strength. The results show that several key characteristics of active regions can be reproduced simultaneously by such magnetic flux loops emerging from the interior of the convection zone: the time scale for emergence of an active region, the heliographic latitude range of emergence, and the degree of tilt of active region axes. For all these phenomena, agreement between simulations and observation points to the same value of the field strength: about $10^5$ G. This is also the field strength at which instability is predicted to set in  (Sch\"ussler et al. 1994). It  can thus be identified with Babcock's (then still unquantified) critical field strength. 

Within this picture, a further piece of evidence that would otherwise be a disconnected observation  finds its natural place. After formation of a spot, its position drifts a bit in latitude and longitude (`proper motion'), with a random component superimposed on a systematic drift. The random component varies quasiperiodically on a time scale of a few days, decaying with time (e.g. Herdiwijaya et al.1997). For a field strength of $10^5$ G at the base, the Alfv\'en travel time along the flux strand from the base to the surface is around 3 days. The random proper motions  of spots are thus neatly interpreted as reflecting the `settling' of a sunspot to its equilibrium position after the eruption process is completed. \bhl [In addition to this random component there is a systematic drift in longitude, corresponding to the increasing separation between the two polarities. This was explained as due to the tension in the sub-surface magnetic field by van Ballegooijen (1982), and reproduced in simulations of rising flux tubes (Caligari et al. 1995)].\ehl

\subsection{Assessment of the turbulent convective dynamo view}
The success of the rising flux tube simulations was not immediately seen as a threat to mean field models. In line with the status of active region phenomenology in mean field models, the eruption process would simply be of marginal significance: it would just be some secondary manifestation of the mean field dynamo operating in deeper layers. The success of the simulations, if it is not accidental, has ominous consequences, however, since the agreement with each of the observations only holds if the field strength in the deep interior of the convection zone is around $10^5$ G. The energy density in such a field is at least two orders of magnitude larger than the kinetic energy of the convective turbulence invoked in mean field models. If taken serious as a diagnostic, the observed mode of emergence of active regions thus implies that the third prediction also fails. 

The situation is more serious, however, since at this field strength the rising flux tubes are so strong that convective turbulence can have little effect on them. The positions of active regions on the surface must consequently correspond reasonably with those of their anchors at the base, also explaining the regularity with which active regions follow Hale's and Joy's laws, \bhl and the proper motions of sunspots mentioned above\ehl. Active regions as seen at the surface are therefore not a manifestation of a convective mean field dynamo, even if it were to exist somewhere in the convection zone. This is a major setback for this theory, since observations of active regions are the dominant source of information we have about the solar cycle, and virtually the only source ever used for mean field parameter fitting.

Considering these spectacular failures of the turbulent mean field dynamo para\-digm for the solar cycle it is useful to reflect for a moment how it could have survived for several decades, and is still going strong (cf.\ R\"udiger and Hollerbach 2004, Brandenburg 2005, 2009, Tobias 2005, Jones et al. 2009, Charbonneau 2005). Major problems like the lack of a stable theoretical foundation for the equations used, and the lack of connection with most of the relevant observations should have been reasons to pause and reflect on the basis of the enterprise. Instead such problems, when faced at all, were usually countered with these arguments: 

i) in a complicated problem one has to start somewhere,

ii) the key to a dynamo cycle is the recreation of a poloidal component, and this is included in the mean field dynamo equation.

The first argument was perfectly reasonable at the time of the formulation of mean field electrodynamics, (c.f. Parker 1955, Steenbeck et al. 1966), but half a century hence it is beginning to wear a bit thin.

The second argument is essentially a semantic kludge. At a sufficiently abstract level, the nature of the solar cycle as a combination of winding up and an alpha-effect need not be contested. What should have been considered more critically, however, is the question how much more the use of mean field dynamo formalisms provides, above just the alpha-effect that was already put in as an assumption from the beginning. The applied mathematical attractions of solving this equation in countless variations of geometry and parameters appears to have led to a misplaced sense of reality. The failure of the theory to show demonstrable progress by providing increasing contact with observations should have been a warning here. The mean field dynamo model for the solar cycle is best regarded as a mirage. Remarkably, it still keeps a substantial community busy searching for an oasis.

\bhl  The theoretical basis of mean field electrodynamics has always been problematic. The series expansions used to derive mean field equations, for example are known to diverge unconditionally at large Reynolds numbers\ehl. In the mean time, it is now also being called into question by the results from high resolution numerical simulations of magnetic fields driven by imposed small scale forces. From these simulations it is becoming clear that large scale fields do not appear from small scales as expected  (e.g. Cattaneo and Hughes, 2010), at least not under the generic conditions where the mean field dynamo equation was applied. The occurrence of {\em small scale} dynamos (i.e. the exponential growth of magnetic energy on small scales) in the interaction of a magnetic field with turbulence is now also somewhat in doubt. \bhl Standard wisdom (the `proper' view, Brandenburg 2010), that  an appropriately complex velocity field is sufficient to produce a selfsustained small-scale magnetic field turns out to be incorrect.  Selfsustained fields have been found in such flows when the viscosity is larger than the magnetic diffusivity (magnetic Prandtl number P$_{\rm m}>1$, e.g. Schekochihin et al. 2005).  For the case P$_{\rm m}<1$ however, conflicting  results are reported on the presence or absence of selfsustaining fields, depending also on numerical method.  In the case P$_{\rm m}>1$, the actual operation mechanism of the field amplification does  not agree with conventional ideas based on cascades in wavenumber space (Schekochihin et al. 2005). Kinematic models of  small-scale dynamo action, i.e. models using an imposed velocity field, which have guided much of previous thinking, do not provide guidance in this context since they are equivalent to assuming an infinite magnetic Prandtl number.\ehl

\section{Tachocline dynamos}

An somewhat older idea to reconcile mean field models with flux emergence observations is that the dynamo works as a turbulent mean field dynamo near the base of the convection zone (Galloway and Weiss 1981, Parker 1993). After its discovery, the narrow shear zone below the convection zone, the tachocline, was quickly identified as a region of choice to operate a mean field dynamo. It contained the strong radial gradient needed in mean field models to produce the drift of active latitudes during the cycle (e.g. Dikpati 2006 and references therein), and required no conceptual adjustments to the models developed before.

This idea, however does not make physical sense. It assumes that the shear zone can be exploited in the same way as the shear between two moving plates in the laboratory. Turbulence generated by the shear exerts stress on both plates, the energy put into the system by the work done against this boundary stress can be tapped to maintain turbulence and a magnetic field.  In the Sun, stresses can be maintained in the convection zone by the rapid momentum exchange due to convective flows.  On the other side, however, in the stable stratification on the interior side of the tachocline, the stress that can be supported by fluid motions is many orders of magnitude weaker, since the stratification is very stable on this side (in terms of the buoyancy $N^2$, the interior is $10^6$ times more stable than the convection zone is unstable). This means that the analogy with shear maintained between two moving plates is incorrect (resembling the Zen exercise to clap with one hand). 

If the analogy is incorrect, what is then the cause of the tachocline? The tachocline is more appropriately treated as a shear flow driven only by the latititudinal dependence of the rotation rate on the convective side, with a free-slip surface on the interior side. The velocities in the tachocline can just be an `imprint' into the interior of the differential rotation with latitude in the convection zone. This has been proposed early on after the discovery of the tachocline by Spiegel and Zahn (1992), who studied the long-term evolution of the rotation in the interior under the effect of a (weak) viscous stress. A more complete analysis of this problem, which stresses the importance of baroclinicity and thermal diffusion, is the `gyrotropic pumping'  model of McIntyre (2007).  In a model by Forg{\'a}cs-Dajka \& Petrovay (2001)  the tachocline is also seen as an imprint of the convection zone on the interior, but this model involves turbulence of unspecified origin inside the tachocline. 

The radial gradient in the tachocline is thus useless for driving a dynamo, since it does not support any significant stress with which a field could be amplified. The latitudinal gradient in the tachocline can of course be tapped, but this is no different from what the bulk of the convection zone can do.  Its shallow imprint into the interior does not add much, and defeats the original idea of using the (radial) tachocline shear to drive a dynamo.

\section{New directions}

\subsection{Compromises}
The mean field paradigm holds that, after ignoring or averaging out most of the surface observations, the bit that is left is still a useful validation of the theory. This point of view is still popular in the literature on the solar cycle.  Another view appears to follow more of an `adiabatic adjustment' approach: attempts are made to incorporate elements like the disregarded observations mentioned in section \ref{obs}, and physics like buoyant instability of the magnetic field into mean field turbulence, as gradual adjustments of the formalism. This view thus attempts to accommodate intrinsically incompatible elements into a mean field approach without questioning its status as the underlying fundamental theory.

The kind of compromises this leads to looks ugly. In one such attempt to conciliate mean field theory with observations, the emergence of strong magnetic flux tubes is in fact acknowledged to account for phenomena observed at the surface. However, the surface fields are then seen as a separate phenomenon, not representative of the real solar cycle. The real cycle takes place, unseen, somewhere in the convection zone in the manner demanded by the mean field dynamo equations (e.g. Tobias 2009). Another proposal (Brandenburg 2005) postulates the existence of a shallow surface layer (a few Mm depth). This layer contains the puzzling observations, again in some form of mean field dynamo, while at the same time shielding the turbulent mean field dynamo happening below it from our view. `Turbulent pumping' is advanced as achieving this. Observations of active region emergence (Fig.\ \ref{trilobite}) that refute such ideas (however vague) are ignored.

\subsection{Weak fields}
A more progressive view within the convective turbulence category is the idea (Durney et al. 1993, Cattaneo, 1999) that turbulent interaction between field and convection is observed at the surface of the Sun in the form of the so-called weak or inner-network fields. These appear as fields of mixed polarity, short life times and low intrinsic strength (Martin 1988). These properties are more in line with a priori expectations about turbulent fields. 

This proposal bypasses the question what causes the strong fields observed as spots and active regions: it only looks at the weak field component and assumes it has a different origin. It is not clear, however, whether the weak fields are really an independent phenomenon: they might just be a part of the solar cycle as seen in active regions. The weak fields might represent either a small scale tail in the distribution of emerging flux units,  or some kind of `debris' from the fragmentation of larger units during the decay phase of active regions. Finally, they might be related to the `annealing step' by which old flux disappears again from the convection zone (section \ref{anneal}). In the latter case, they would be part of the {\em decay} of magnetic fields produced in the cycle, rather than an amplification process. 

\subsection{Numerical simulations}

Convincing 3-D numerical simulations of a solar cycle `from scratch' are likely to remain out reach for the indefinite future. This is because of the well-known problems of dynamical range in length and time scale intrinsic to a convective stellar envelope. A brute-force simulation of the entire convection zone would have to resolve time scales as short as a few seconds in the photospheric layers and above,  as well as years to cover the duration of a cycle. Corresponding length scales range from a kilometer at the surface to a solar radius. Extrapolating Moore's law with a constant doubling time of 1.5 years, the computing resources needed for a simulation at this resolution would become available 100 years from now (Sch\"ussler 2008). Existing `global' simulations of the convection zone or its magnetic cycle are possible only by leaving out key parts of the physics. Usually, the top layers where most of the dynamic range in length- and time scales is located,  are left out. Conclusions drawn from such simulations are unlikely to be very meaningful, \bhl since the easier case of a convective envelope without magnetic fields is already known to produce results that bear no relation to observations when these surface layers are left out from the simulation\ehl.  Individual aspects of the cycle still provide interesting unsolved conceptual problems, however, that may be addressed in isolation before realistic numerical simulations are attempted on more global scales. \bhl Two such problems are discussed in the next subsections\ehl.

\subsection{The annealing step, `turbulent diffusion'}
\label{anneal}
The most challenging problem may well be finding a satisfactory description for the process by which the mass of buoyant vertical flux tubes resulting from a cycle's worth of eruptions gets `annealed' back into a simpler configuration. As the eruption of active regions from the toroidal field proceeds during the cycle, an ever increasing number of magnetic strands develops connecting the surface with the base of the convection zone (cf. Fig.\ \ref{spotsketch}). The sections of field remaining at the base are sufficient to provide the toroidal field of the next cycle, but this picture does not explain how the clutter of strands connecting to the surface gets simplified from one cycle to the next. In dynamo parlance, this is the `turbulent diffusion' step. The difficulty here is that appeal to traditional convective `turbulent diffusion' will not work (even if the concept itself is accepted), since the fields are now much stronger than equipartition with convection (at least near the base of the convection zone where this annealing has to take place). 

At the moment, it is not clear how long this annealing process takes, or even by which mechanism. For a discussion relating to this problem, see Parker (2009). Perhaps the `residence time' of dispersed active regions is substantially longer than the cycle length? In that case a large amount of small scale mixed-polarity magnetic flux should be present distributed over the solar surface. Such a component could remain undetected except at very high spatial resolution. In fact, recent observations of mixed-polarity fields in quiet regions of the solar surface with Hinode and the Swedish 1-m solar telescope appear to show evidence in this direction (cf Pietarila et al. 2009, but see also Deforest et al. 2008) . 

Extrapolating this thought further, it might even be that such a component is relevant for total solar irradiance (TSI, the solar energy flux received at earth), since small scale magnetic fields are known to produce a net brightening of the solar surface (Spruit 1977). If this is the case, the small but systematic decrease of TSI during the current extended minimum, below the previous shorter minima, might be indicative of the decrease of surface magnetic flux in the course of the annealing process.

\subsection{Thermodynamics}

A central question  concerns the thermodynamics of fields of $\sim 10^5$ G at the base of the convection zone. For the field to be wound up quietly over several years before becoming unstable, it has to reside in a stable buoyant equilibrium near the base of the convection zone. In the absence of such a stable equilibrium, the field would rise to the surface on  a time scale of weeks (as it actually does in the emergence of a new active region). Magnetic pressure produces buoyancy, and this has to be compensated for equilibrium to hold. Neutral buoyancy through density equilibrium requires significantly lower temperatures in the field than in its environment. For a field strength of $10^5$ G, the required amount of temperature reduction is some 100 times larger than canonical temperature fluctuations in a mixing length model of convection. 

To solve this equilibrium problem, it is often postulated that flux tubes erupt from stably stratified layers below the base of the convection zone. Though this recognizes the buoyancy problem, it does not actually help much since it begs the question how the magnetic field got to this location in the first place (in particular: on a time scale less than the solar cycle). 

\section{Conclusions}

Observations of active region phenomenology, most of them already old and well-established, show that the solar cycle operates on buoyant instability of the magnetic field itself rather than the conventional view based on interaction with convection. This puts us back to ideas developed half a century ago. Significant steps forward, however are the direct 3-D, radiative numerical MHD simulations which are now beginning to make contact with some of the classical observations. Though these simulations cannot deal with the cycle as a whole, their success in reproducing limited aspects such as the emergence of magnetic flux discussed above, or the observed structure of sunspots (\bhl Heinemann et al. 2007, Scharmer et al. 2008, Rempel et al. 2009\ehl) give confidence for the future. At the same time they clean the table by eliminating a number of dead-end views on the solar cycle, some of which considered well-established thus far. 

At the same time, a number of unsolved questions appear that are specific for the picture of a magnetic cycle operating on buoyant instability. Some of these questions are unlikely to be answered from first principles or numerical simulations. Clues taken from observations may well play an important role in making progress in figuring out the physics relevant for these questions. As the history of the subject shows, however, taking observational clues serious will require one to jettison the turbulent mean field baggage that has impeded the development of a sensible theory of the solar cycle for so long. This process would be assisted by healthy skepticism on the part of the observational community. In fact, it is rather surprising how easily observers have acquiesced in the past to the treatment of their data by mean field theories (`sorry but your observations are just turbulence, they have to be averaged out'). 

\medskip

\section{References}
\parindent=-10pt

\bibitem{} Altrock, R.~C.\ (2010), arXiv:1002.2401

\bibitem{}Babcock, H.W. (1961). \apj \textbf{133}, 1049.

\bibitem{}Babcock, H.W. (1963). \textit{Ann. Rev. Astron. Astrophys.} \textbf{1}, 41. 

\bibitem{}Balbus, S.~A., \& Hawley, J.~F.\ (1991). \apj \textbf{376}, 214.

\bibitem{}Brandenburg, A. (2005). \apj \textbf{625}, 539.

\bibitem{} Brandenburg, A.\ (2009), Plasma Physics and Controlled Fusion \textbf{51}, 124043 

\bibitem{} Brandenburg, A.\ (2010), \mnras \textbf{401}, 347 

\bibitem{}Brandenburg, A.,  Kleeorin, N., \& Rogachevskii, I.\ (2010), Astronomische Nachrichten \textbf{331}, 5 

\bibitem{} Caligari, P., Moreno-Insertis, F., \& Schussler, M.\ (1995), \apj \textbf{441}, 886 

\bibitem{} Cattaneo, F.\ (1999), \apjl \textbf{515}, L39 

\bibitem{} Cattaneo, F., \& Hughes, D.~W.\ (2009), \mnras \textbf{395}, L48 

\bibitem{} Charbonneau, P. 2005, \textit{Living Reviews in solar phsyics}, \hfil\break http://solarphysics.livingreviews.org/Articles/lrsp-2005-2/

\bibitem{} Cheung, M.~C.~M., Sch{\"u}ssler, M., Tarbell, T.~D., \& Title, A.~M.\ (2008), \apj \textbf{687}, 1373 

\bibitem{}Cline, K.~S., Brummell, N.~H., \& Cattaneo, F. ( 2003). \apj \textbf{599}, 1449.

\bibitem{}Cowling, T.G. (1953), in \textit{The Sun}, G. Kuiper, ed., Univ. of Chicago Press, Chapter 8 

\bibitem{} Deforest, C.~E., Lamb, D.~A., Berger, T., Hagenaar, H., Parnell, C., \& Welsch, B.\ 2008, AGU Spring Meeting Abstracts, 1 

\bibitem{} Dikpati, M.\ (2006), Advances in Space Research, \textbf{38}, 839 

\bibitem{}D'Silva, S., \& Choudhuri, A.~R. (1993). \aap \textbf{272}, 621. 

\bibitem{}Durney, B.~R., De Young, D.~S., \& Roxburgh, I.~W. (1993). \solphys \textbf{145}, 207.

\bibitem{}Elsasser, W.~M. (1956). \textit{Reviews of Modern Physics} \textbf{28}, 135 

\bibitem{}Fan, Y., Fisher, G.~H., \& McClymont, A.~N. (1994). \apj \textbf{436}, 907

\bibitem{} Forg{\'a}cs-Dajka, E., \& Petrovay, K.\ (2001), \solphys, 203, 195 

\bibitem{}Galloway, D.~J., \& Weiss, N.~O. (1981). \apj \textbf{243}, 945 

\bibitem{}Gilman, P.~A. (2005). Astronomische Nachrichten \textbf{326}, 208 

\bibitem{}Hawley J.F., Gammie C.F., Balbus S.A. (1996). \apj \textbf{464}, 690 

\bibitem{} Heinemann, T., Nordlund, {\AA}., Scharmer, G.~B., \& Spruit, H.~C.\ (2007), \apj  \textbf{ 669}, 1390 

\bibitem{} Herdiwijaya, D., Makita, M., \& Anwar, B.\ (1997), \pasj \textbf{49}, 235 

\bibitem{} Howe, R.\ (2009), Living Reviews in Solar Physics, \textbf{6}, 1, arXiv:0902.2406

\bibitem{} Jones, C.~A., Thompson, M.~J., \& Tobias, S.~M.\ (2009), Space Science Reviews, \textbf{114} , 	
DOI: 10.1007/s11214-009-9579-5

\bibitem{}Kitiashvili, I.N., Kosovichev, A.G., Wray, A.A., Mansour, N.N.\ (2010), arXiv:1004.2288v1 

\bibitem{} Kleeorin, N., \& Rogachevskii, I.\ (2008), \pre \textbf{77}, 036307 

\bibitem{}Leighton, R.~B. (1969). \apj \textbf{156}, 1 

\bibitem{}Leroy, J.L., and Trellis, M. (1974) \aap \textbf{35}, 283

\bibitem{}Martin, S.~F. (1988). \solphys \textbf{117}, 243. 

\bibitem{}Mathis, S., \& Zahn, J.-P. (2004). \aap \textbf{425}, 229


\bibitem{} McIntyre, M.~E.\ (2007), in The Solar Tachocline, eds. D.W.\ Hughes et al., CUP, p183 

\bibitem{}Meyer, F., Schmidt, H.~U., Wilson, P.~R., \& Weiss, N.~O. (1974) \mnras \textbf{169}, 35.

\bibitem{}Moreno-Insertis, F. (1986). \aap \textbf{166}, 291.


\bibitem{}Parker, E.N. (1955). \apj \textbf{122}, 193.

\bibitem{}Parker, E.N. (1979). \textit{Cosmical Magnetic Fields} (Clarendon Press, Oxford).

\bibitem{} Parker, E.~N. (1993). \apj  \textbf{408}, 707 

\bibitem{} Parker, E.~N.\ (2009), Space Science Reviews, 144, 15 

\bibitem{} Pietarila Graham, J., Danilovic, S., \& Sch{\"u}ssler, M.\ 2009, \apj, 693, 1728 

\bibitem{}Pitts E., Tayler R.J. (1986). \mnras \textbf{216}, 139. 

\bibitem{}Prendergast, K.H. (1956). \apj \textbf{123}, 498.

\bibitem{}Proctor, M.~R.~E., \& Weiss, N.~O. (1982). \textit{Reports of Progress in Physics} \textbf{45}, 1317.

\bibitem{} Rempel, M., Sch{\"u}ssler, M., Cameron, R.~H., \& Kn{\"o}lker, M.\ 2009, Science, \textbf{325}, 171 

\bibitem{}R\"udiger, G., \& Hollerbach, R. (2004). \textit{The magnetic universe : geophysical and astrophysical dynamo theory}. Weinheim: Wiley-VCH

\bibitem{} Scharmer, G.~B., Nordlund, {\AA}., \& Heinemann, T.\ (2008), \apjl \textbf{677}, L149 

\bibitem{}Sch\"ussler, M., Caligari, P., Ferriz-Mas, A., \& Moreno-Insertis, F. (1994). \aap \textbf{281}, L69. 

\bibitem{} Schekochihin, A.~A., Cowley, S.~C., Taylor, S.~F., Maron, J.~L., \& McWilliams, J.~C.\ (2004), \apj, \textbf{612}, 276 

\bibitem{} Sch{\"u}ssler, M.\ (2008), 12th European Solar Physics Meeting, Freiburg, Germany, held September, 8-12, 2008.~Online at http://espm.kis.uni-freiburg.de/, p.1.1, 12, 1 

 \bibitem{} Spiegel, E.A. \& Zahn, J.-P. (1992). \aap \textbf{265}, 106.

\bibitem{}Spruit, H.~C. (1977), \solphys \textbf{55}, 3.

\bibitem{}Spruit, H.~C.  and Roberts, B.\ (1983)  \textit{Nature} \textbf{304}, 401 

\bibitem{}Spruit, H.~C. (2002). \aap \textbf{381}, 923.

\bibitem{} Steenbeck, M., Krause, F., R\"adler, K.-H.\ (1966), Zeitschrift f\"ur Naturforschung Teil A, \textbf{21}, 369 

\bibitem{} Strous, L.~H., Scharmer, G., Tarbell, T.~D., Title, A.~M., \& Zwaan, C.\ 1996, \aap \textbf{306}, 947 

\bibitem{}Tarbell, T., Ferguson, S., Frank, Z., Shine, R., Title, A., Topka, K., \& Scharmer, G. (1990). \textit{IAU Symposium} \textbf{138}, 147 

\bibitem{} Tobias, S.\ (2005), Advances in astronomy, ed. J. M. T. Thompson. \textit{Royal Society Series on Advances in Science}, Vol. 1. London: Imperial College Press, 355 

\bibitem{} Tobias, S.~M.\ (2009), \textit{Space Science Reviews} \textbf{144}, 77 

\bibitem{} van Ballegooijen, A.~A.\ (1982), \aap \textbf{113}, 99 

\bibitem{}Vrabec, D.\ (1974), in R.G. Athay, ed., Chromospheric Fine Structure, IAU Symp. 56, p221

\bibitem{}Weiss, N.~O. (1964). \mnras \textbf{128}, 225. 

\bibitem{}Weiss, N.~O. (1981). JGR \textbf{86}, 11689. 

\bibitem{}Weiss, N.~O. (1989). ASSL \textbf{156}: \textit{Accretion Disks and Magnetic Fields in Astrophysics}, p.11 

\bibitem{}Weiss, N.~O. (1993). ASSL \textbf{183}: \textit{Physics of Solar and Stellar Coronae}, p.541 

\bibitem{}Weiss, N.~O. (1994). in \textit{Lectures on Solar and Planetary Dynamos}, eds. M.R.E.\ Proctor and A.D.\ Gilbert. ISBN 0 521 46142 1 and ISBN 0 521 46704 7, Cambridge University Press, p.59

\bibitem{}Weiss, N.~O. (1997). \textit{Past and present variability of the solar-terrestrial system}, "Enrico Fermi" : course CXXXIII, eds. G Castagnoli, A. Provenzale, Oxford: IOS, p.325 




\bibitem{}Zwaan, C.\ (1978), \solphys \textbf{60}, 213.
 

\end{document}